\documentclass[preprint,12pt]{elsarticle}

\usepackage{graphicx}
\usepackage{amssymb}

\usepackage{amsthm}

\usepackage{setspace}
\usepackage[margin=1.5in]{geometry}    
\usepackage[utf8]{inputenc}
\usepackage{algorithm}
\usepackage[noend]{algpseudocode}

\makeatletter
\def\BState{\State\hskip-\ALG@thistlm}
\makeatother
\usepackage{enumitem} 
\usepackage{tablefootnote}
\usepackage{caption}
\usepackage{graphicx}
\usepackage{booktabs}
\usepackage{amsfonts}
\usepackage{amsthm}
\usepackage{amsmath,bm}
\usepackage{amssymb}
\usepackage{mathtools}
\usepackage{subfig}
\usepackage{multirow}
\usepackage{bbm}
\usepackage{hyperref}

\usepackage{array}

\usepackage[colorinlistoftodos]{todonotes}
\usepackage{comment}

\usepackage{mathrsfs}

\usepackage{epstopdf}
\usepackage{enumitem}  

\journal{Transportation Research Part C}

\begin{document}

\begin{frontmatter}


\title{A GRU-based Mixture Density Network for Data-Driven Dynamic Stochastic Programming}



\author[1,3,4]{Xiaoming Li}
\author[1]{Chun Wang}
\author[2]{Xiao Huang}
\author[4]{Yimin Nie}

\address[1]{Department of Information and System Engineering, Concordia University, Montreal, QC, Canada}
\address[2]{John Molson School of Business, Concordia University, Montreal, QC, Canada}

\address[3]{School of Computer, Shenyang Aerospace University, Shenyang, Liaoning, P.R. China}
\address[4]{Ericsson INC. Global Artificial Intelligence Accelerator (GAIA) innovation hub, Montreal, QC, Canada}

\begin{abstract}
The conventional deep learning approaches for solving time-series problem such as long-short term memory (LSTM) and gated recurrent unit (GRU) both consider the time-series data sequence as the input with one single unit as the output (predicted time-series result). Those deep learning approaches have made tremendous success in many time-series related problems, however, this cannot be applied in data-driven stochastic programming problems since the output of either LSTM or GRU is a scalar rather than probability distribution which is required by stochastic programming model. To fill the gap, in this work, we propose an innovative data-driven dynamic stochastic programming (DD-DSP) framework for time-series decision-making problem, which involves three components: GRU, Gaussian Mixture Model (GMM) and SP. Specifically, we devise the deep neural network that integrates GRU and GMM which is called GRU-based Mixture Density Network (MDN), where GRU is used to predict the time-series outcomes based on the recent historical data, and GMM is used to extract the corresponding probability distribution of predicted outcomes, then the results will be input as the parameters for SP. To validate our approach, we apply the framework on the car-sharing relocation problem. The experiment validations show that our framework is superior to data-driven optimization based on LSTM with the vehicle average moving lower than LSTM.
\end{abstract}

\begin{keyword}
Data-Driven Optimization (DDO)\sep Stochastic Programming (SP)\sep Gated Recurrent Unit (GRU)\sep Gaussian Mixture Model (GMM)\sep Mixture Density Network (MDN)\sep  
\end{keyword}

\end{frontmatter}


\setlength{\baselineskip}{20pt}

\section{Introduction}
A variety of deep learning (DL) models and algorithms have been proposed and successfully solved a great many of applications fields such as computer vision, natural language processing, data analysis etc. Due to its great success, recently, leveraging machine learning (ML) and deep learning approaches to support decision-making problem has attracted huge attentions in operations research community. Unlike regular ML approaches, DL methods are capable of dealing with intrinsic and potential features that are hidden behind the complex data. In data-driven optimization frameworks, the uncertainty is modeled based on complex data which may has great impact on the optimization solutions. The inaccurate parameters that are derived from complex data may lead the optimizations model sub-optimal or even infeasible. In this sense, data-driven optimization could be benefit from utilizing DL tools.

Recently, decision-making under uncertainty has been applied in various fields such as intelligent transportation, network optimization, scheduling problems, supply chain management etc. In this work, we focus on stochastic programming (SP) technique which is aiming to find the optimal solution that maximize / minimize the expected value of objective function while satisfying all the scenarios that are obtained from uncertain parameters. Conventionally, SP assumes that the probability distribution of uncertain parameters is known from perfect knowledge. In reality, however, it is difficult even impossible to obtain the accurate probability distribution from complex data. It is worth noting that the inaccurate probability distribution may lead the optimization solution to be sub-optimal even infeasible, therefore, it is quite necessary to integrate ML / DL approaches to improve the solution quality of SP. 

In this paper, we consider using SP to solve time-series decision-making problems, which has been extensively studied and widely applied in intelligent transportation domain. As discussed in the previous section, probability distribution is required by SP, however, the existing deep learning approaches for time-series predication such as long-short term memory (LSTM)\cite{hochreiter1997long} and gated recurrent unit (GRU)\cite{cho2014learning, chung2014empirical}, both of them return single unit (scalar) output as the predicted results which cannot be used as the parameters for SP. To fill the gap, we devise an innovative deep neural network which involves GRU and mixture density network (MDN) \cite{bishop1994mixture} called GRU-MDN for the time-series probability distribution prediction. Further, we propose a novel data-driven dynamic stochastic programming framework that integrates GRU-MDN along with SP to solve time-series decision-making problems under uncertainty. To best of our knowledge, this is the first work that combine DL and SP for time-series decision-making problems.

The contributions in this work are two-folded: (1) a novel GRU-MDN deep neural network is devised to predict probability distribution of time-series data, (2) stochastic programming is seamlessly integrated with GRU-MDN to formulate relevant problems. The remainder of this paper is organized as follows, the details of data-driven dynamic stochastic programming framework is discussed in section \ref{S:2}, next we apply the framework using a toy example which is in section \ref{S:3}, the conclusions and future work is summarized in section \ref{S:4}. 
\section{GRU-MDN stochastic programming framework} \label{S:2}
To make data-driven SP that is capable of solving time-series problem, we propose a GRU-base mixture density network called GRU-MDN. The framework involves three components, GRU is in charge of predicting customer demands, GMM focuses on the probability distribution that are based on the outcomes from GRU, the SP is in charge of modeling uncertainty. We will elaborate our framework in this section.

\subsection{Gated Recurrent Unit}

\begin{figure}
    \centering
    \includegraphics[scale = 0.3]{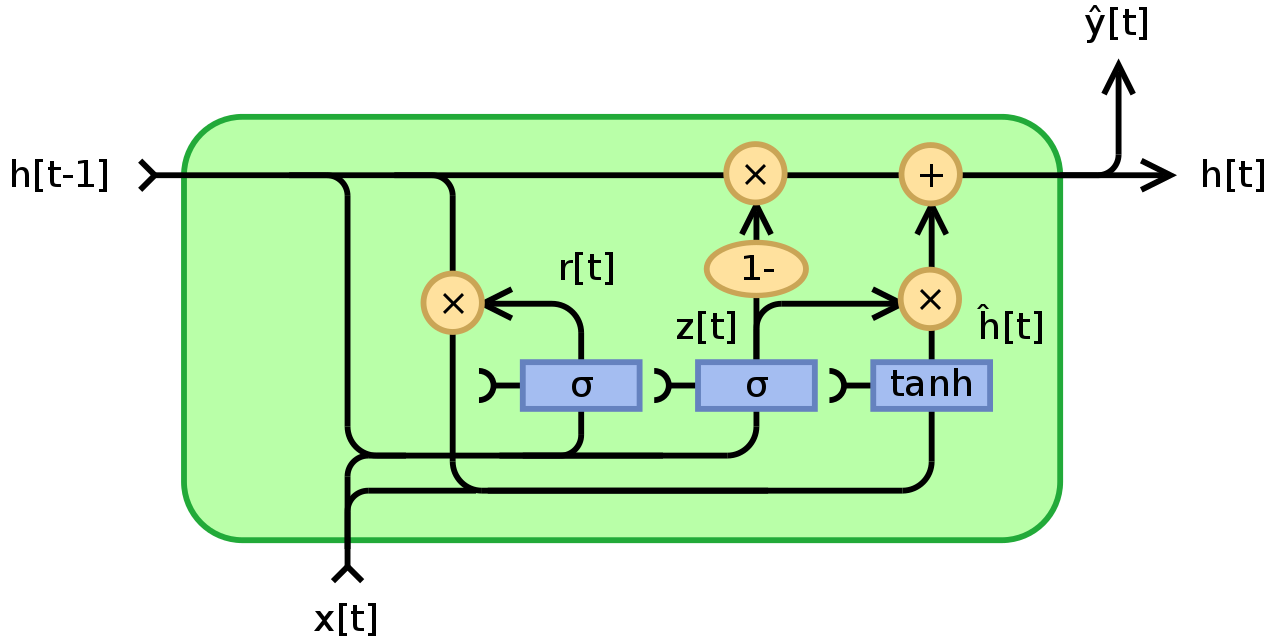}
    \caption{Gated Recurrent Unit}
    \label{fig:gru}
\end{figure}

Unlike the single and simple building block in RNNs, LSTM uses forgetting and gating mechanisms to select and filter information that is necessary for future computation. Figure \ref{fig:gru} shows the detail gating mechanism of GRU.

The state of GRU for each time step $t$ is given by the following equations.
 
\begin{equation}
\begin{array}{l}
z_{t}=\sigma_{g}\left(W_{z} x_{t}+U_{z} h_{t-1}+b_{z}\right) \\
r_{t}=\sigma_{g}\left(W_{r} x_{t}+U_{r} h_{t-1}+b_{r}\right) \\
h_{t}=z_{t} \odot h_{t-1}+\left(1-z_{t}\right) \odot \phi_{h}\left(W_{h} x_{t}+U_{h}\left(r_{t} \odot h_{t-1}\right)+b_{h}\right)
\end{array}
\end{equation}
where $x_{t}$ is the input vector, $h_{t}$ is the output vecotr, $z_{t}$ is the update gate vector, $r_{t}$ is the reset gate vector, $W,U,b$ are the parameter matrices and vector. $\sigma_{g}$ and $\phi_{h}$ are activation functions in sigmoid and hyperbolic types, respectively.

\subsection{Mixture Density Network}
MDN is a variant of a neural network whose output is probability distribution(s) rather than single unit for most of neural networks. The basic idea of MDN is \textit{combining a deep neural network (DNN) and a (group of) mixture of distributions}.  Actually, most of the modern DNN architectures such as CNN, RNN and LSTM can be extended to become the special MDNs. The DNN provides the parameters for multiple distributions, which are then mixed by some weights. Also, These weights are provided by the DNN. Notice that the true distribution of the input data can be any type which cannot be described by the parametric methods (e.g. Gaussian distribution, Poisson distribution). Actually, the non-parametric method such as kernel density estimation (KDE) is able to handle with the arbitrary probability distribution learning problem. However, KDE method cannot be directly applied for probability distribution storage in MDN since KDE is a non-parametric approach which implies that it may contain infinite parameters that cannot be store by finite number of neurons. Therefore, the semi-parametric approach - Gaussian mixture model (GMM) is adopted to overcome the problem in MDN, which is formulated as follows.

$$p(\boldsymbol{X \big| \theta})=\sum_{i=1}^{K} w_{i} \mathcal{N}\left(\boldsymbol{X \big| \mu}_{i}, \boldsymbol{\Sigma}_{i}\right)$$
where $\theta = (W, \mu, \Sigma)$, $W = (w_{1}, w_{2}, \cdots, w_{K})$, $\mu = (\mu_{1}, \mu_{2}, \cdots, \mu_{K})$ and $\Sigma = (\Sigma_{1}, \Sigma_{2}, \cdots, \Sigma_{K})$. $K$ is the number of Gaussian distributions. Generally, GMM can be considered as a group of Gaussian distributions with different weights, where the $i-th$ Gaussian is determined by weight $w_{i}$, means $\mu_{i}$ and covariance matrix $\Sigma_{i}$ (variance for univariate Gaussian). Then the predicted probability distribution can be represented using GMM by adjusting the parameter $\theta$. In this work, we use expectation maximization (EM)\cite{dempster1977maximum} algorithm to determine the parameter of GMM, which is summarized as follows.

\begin{algorithm}
\caption{EM}\label{euclid}
\label{al}
\hspace*{\algorithmicindent} 
\textbf{Input: The GMM $\sum_{k=1}^{K} \pi_{k} \mathcal{N}\left(\boldsymbol{x} | \boldsymbol{\mu}_{k}, \mathbf{\sigma}_{k}\right)$}  
\\
\hspace*{\algorithmicindent} 
\textbf{Output: GMM}

\begin{algorithmic}[1]

\State Initialize $\mu_{j}, \Sigma_{j}$ and $\pi_{j}$, $j=1, \cdots, K$
\State E-step. Compute $$
\gamma_{n j}=\frac{\pi_{j} \mathcal{N}\left(\xi_{n} | \mu_{j}, \Sigma_{j}\right)}{\sum_{i=1}^{K} \pi_{i} \mathcal{N}\left(\xi_{n} | \mu_{i}, \Sigma_{i}\right)}
$$
\State M-step. Re-estimate $$
\mu_{j}^{\mathrm{new}}=\frac{1}{N_{j}} \sum_{n=1}^{N} \gamma_{n j} \xi_{n}, \Sigma_{j}^{\mathrm{new}}=\frac{1}{N_{j}} \sum_{n=1}^{N} \gamma_{n j}\left(\xi_{n}-\mu_{j}^{\mathrm{new}}\right)\left(\xi_{n}-\mu_{j}^{\mathrm{new}}\right)^{\top}, \pi_{j}^{\mathrm{new}}=\frac{N_{j}}{N}
$$
where 
$$
N_{j}=\sum_{n=1}^{N} \gamma_{n j}
$$
\State Check whether the convergence is satisfied. If not, return to step 2.

\end{algorithmic}
\end{algorithm}

\begin{figure}
    \centering
    \includegraphics[scale = 0.2]{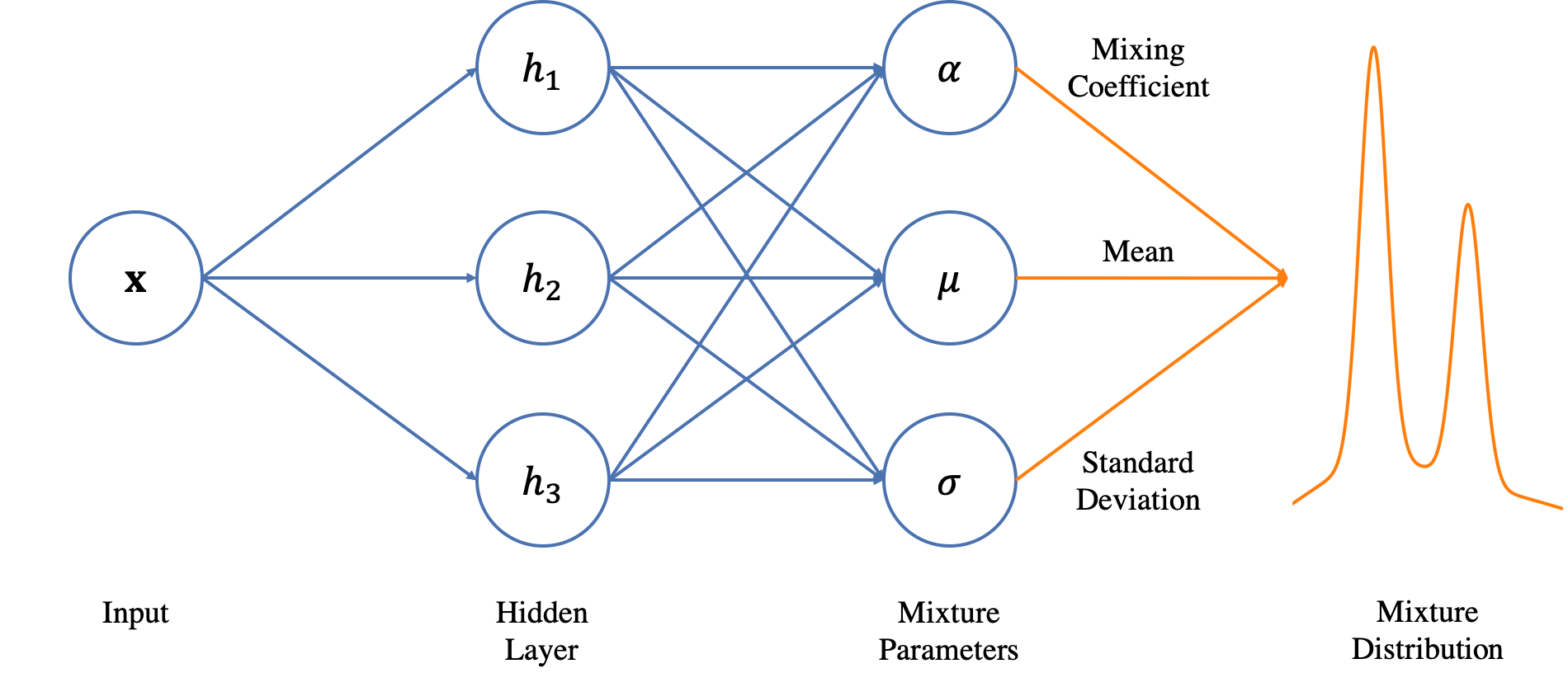}
    \caption{Architecture of Mixture Density Network}
    \label{fig:mdn}
\end{figure}

The structure of MDN can be shown in the figure\ref{fig:mdn}.

\begin{figure}
    \centering
    \includegraphics[scale=0.4]{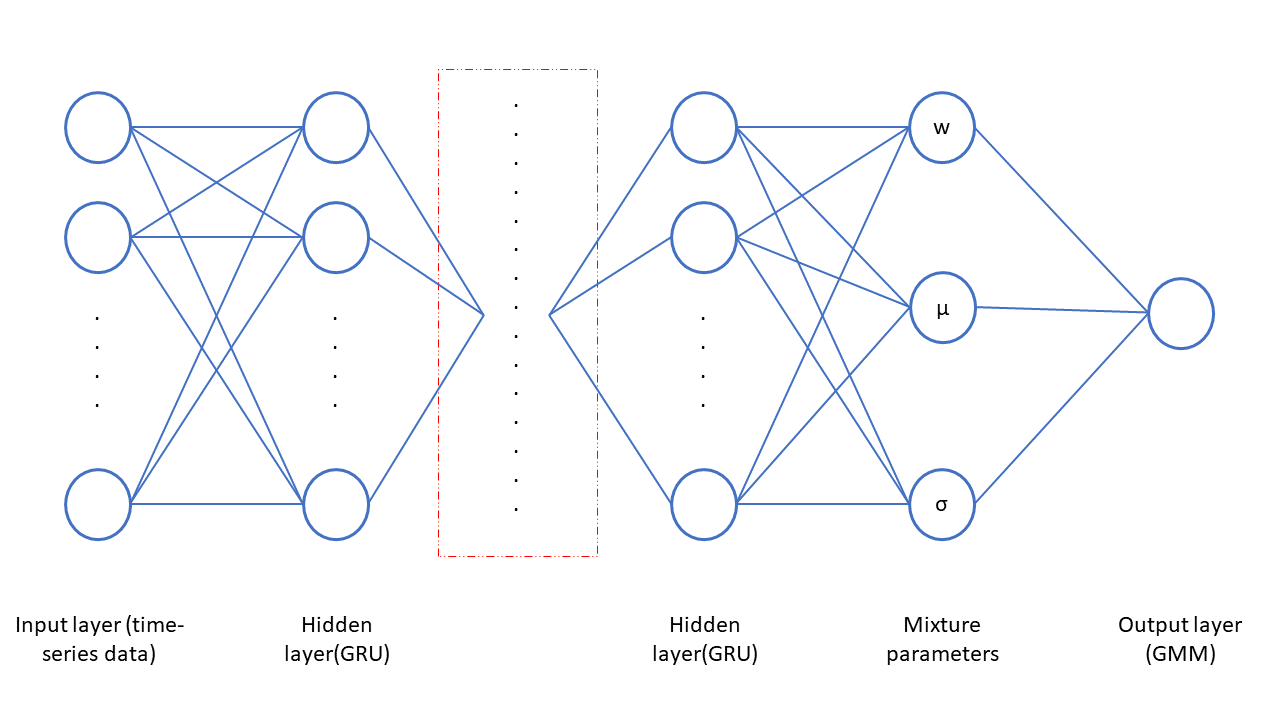}
    \caption{The Structure of GRU-MDN}
    \label{fig:GRU-MDN}
\end{figure}

One of the significant contribution of this work is to combine GRU and MDN, our proposed architecture of GRU-MDN can be shown in figure \ref{fig:GRU-MDN}.

\subsection{Data-driven dynamic stochastic programming framework}
\begin{figure}
    \centering
    \includegraphics[scale = 0.4]{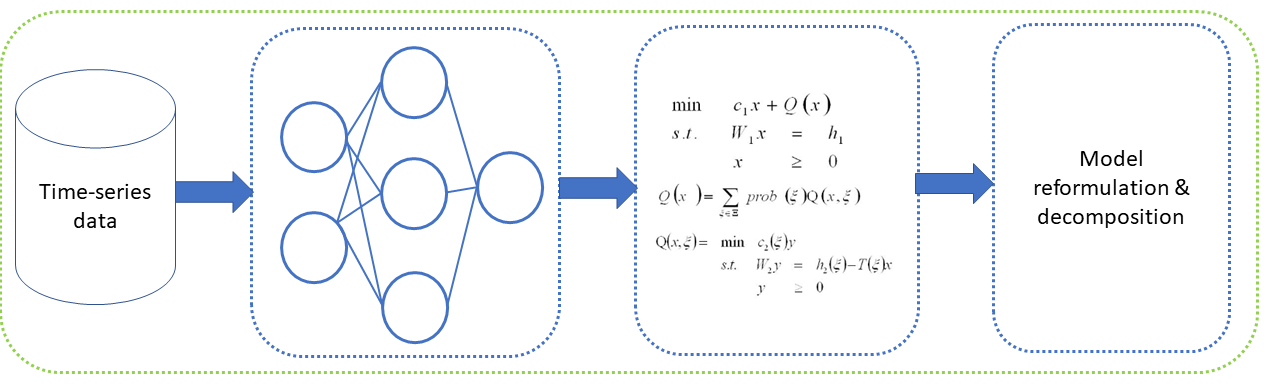}
    \caption{GRU-MDN SP Framework}
    \label{fig:framework}
\end{figure}

Finally, we come to summarize our data-driven SP framework which is displayed in the figure \ref{fig:framework}.

There are four major components involving in the proposed framework. Specifically, the time-series data is the input of the framework, then the predicted time-series probability distribution is obtained via GUR-MDN, the distribution is input as the parameters of SP, after that various model reformulation and decomposition algorithms can be applied to solve the SP model. 
\section{Case Study}\label{S:3}

To validate our data-driven dynamic SP framework, we investigate the car-sharing relocation problem (CSRP) which is referred in \cite{li2020ddksp}. Specifically, we will use the same stochastic programming model and data sets.

\subsection{Experimental Setting}
We use real data from New York taxi trip record data set from January 2017 to June 2019\footnote{https://www1.nyc.gov/site/tlc/about/tlc-trip-record-data.page}. The entire data set is split into training set (from January 2017 to March 2019), and testing set (from April 2019 to June 2019). We compare our proposed approach with data-driven deterministic optimization model where the demand is obtained from the typical time-series prediction approach - LSTM.

The GRU-MDN models are implemented using Python 3.7 + tensorflow 2.1 under the platform CUDA 10.2 GPU, 16GB RAM, Ubuntu 18.04, the mathematical models are solved by Gurobi \footnote{https://www.gurobi.com/downloads/gurobi-software/} 9.0 academic version using Python 3.7 under the platform Intel i7 CPU, 32GB RAM, Windows 10.

\subsection{Experimental Validation}

In order to compare the optimization performance of GRU-MDN and LSTM. We select the similar deep neural network stricture which is shown in Table \ref{table:structure of deep neural network}.

\begin{table}[]
\centering
\caption{The structures of two deep neural network}
\begin{tabular}{|c|c|c|c|}
\hline
         & Input Layer & Hidden Layers & Output Layer \\ \hline
GRU-MDN & ws = 10     & (256, 128)    & (3*3, 1)     \\ \hline
LSTM     & ws = 10     & (256, 128)    & 1            \\ \hline
\end{tabular}
\label{table:structure of deep neural network}
\end{table}
We select the window size equals 10 as the input layer for both GRU-MDN and LSTM. There are two hidden layers for each neural network with the number of neurons 256 and 128, respectively. Since GRU-MDN returns a probability distribution, we use GMM as the output which involves 3 Gaussian.

It is worth noting that although the structures of both deep neural network are quite similar, the output of GRU-MDN is a predicted time-series distribution, which integrates a two-stage stochastic programming model, while the output of LSTM is a predicted time-series value, which integrates a deterministic model. Then we use the first-stage solutions that are obtained from GRU-MDN and LSTM to test on the testing set (91 days from April 1, 2019 to June 30, 2019). The comparison of the experiment results is shown in Table \ref{table:GRU-MDN vs. LSTM}.

\begin{table}[]
\centering
\caption{Comparison between GRU-MDN Stochastic Programming Model and LSTM Deterministic Model}
\begin{tabular}{|c|c|c|c|}
\hline
         & Average Revenue & Average Cost & Average Moving \\ \hline
GRU-MDN & \$947552.6        & \$415601.5     & 231.4615       \\ \hline
LSTM     & \$921922.4        & \$438014.2     & 248.7143       \\ \hline
\end{tabular}
\label{table:GRU-MDN vs. LSTM}
\end{table}

The experiment results show that GRU-MDN with SP model is able to yield more average revenue with relatively lower average cost compared to LSTM with deterministic model. Additionally, the moving average of GRU-MDN is 6.94\% lower than LSTM.
\section{Concluding Remarks}\label{S:4}
In this work, we developed a practical data-driven dynamic stochastic programming framework for time-series problem. The approach integrates a GRU-MDN deep neural network along with a two-stage stochastic programming model. Our proposed methodology provides a very efficient framework for data-driven dynamic SP technique. Furthermore, the framework does not apply in the discussed example only. Actually, as a potential extension, the framework can be applied in a number of different applications by replacing the components. For instance, the component of SP can be replaced by distributionally robust optimization \cite{delage2010distributionally} (DRO) which relies on the ambiguity set that contains a family of probability distributions. We believe that the framework is capable of dealing with DRO modeling problems by minor modifications on GRU-MDN. Additionally, different model decomposition algorithms such as L-shape, column generation can be adopted for model solving according to the characteristics of mathematical models.

Although the proposed framework utilize the historical data to solve the time-series decision-making under uncertainty, it does not consider the prior probability distribution which may be quite informative for SP. We believe that using prior probability distribution information may improve the SP solution in a very effective way, therefore, we will investigate  Bayesian learning with SP for data-driven SP framework in our future work.  





\bibliographystyle{model1-num-names}
\bibliography{sample.bib}







\end{document}